# Soliton Staircases and Standing Strain Waves in Confined Colloidal Crystals

Yu-Hang Chui[1*], Surajit Sengupta[2] and Kurt Binder[1]

[1] Institute of Physics, Johannes-Gutenberg University

D-55099 Mainz, Staudinger Weg 7, Germany

[2] S.N. Bose National Centre for Basic Sciences, Block JD, Sector III,

Salt Lake, Kolkata 70098, India

* Corresponding author, email address: yuhang@graduate.hku.hk

**Abstract**

We show by computer simulation of a two-dimensional crystal confined by corrugated walls that confinement can be used to impose a controllable mesoscopic superstructure of predominantly mechanical elastic character. Due to an interplay of the particle density of the system and the width $D$ of the confining channel, ``soliton staircases'' can be created along both parallel confining boundaries, that give rise to standing strain waves in the entire crystal. The periodicity of these waves is of the same order as $D$. This mechanism should be useful for structure formation in the self-assembly of various nanoscopic materials.

**Introduction**

Colloidal particles and their self-assembly can be influenced by laser fields [1], confinement in photolithographically fabricated channels [2], and various other external fields [3]. Also the interparticle interaction is controllable (e.g. in magneto-rheological fluids via a magnetic field [4]). Due to their visibility in optical microscopes where confocal microscopy allows the tracking of individual particles in real time [5] colloidal particles and the crystals they form are ideal model systems for basic studies in the physics of condensed matter and materials science [6,7]. Such systems have widespread applications, e.g. efficient DNA separation [8].

Thus the present work also deals with a model [9] for a confined colloidal crystal in two dimensions, for which we consider the case of increasing misfit between the distance between the two corrugated walls and the lattice structure that the system would adopt at the chosen density in a stress-free crystal. With increasing misfit (i.e. strain) we observe that the stress increases up to some critical value, where a transition occurs that reduces the number of crystalline rows parallel to the boundaries by one. At constant density, the extra particles of the row that disappears are distributed in the system such that along the walls a ``soliton staircase'' [10] is created, accompanied by a pattern of standing strain waves in the crystal.

This new type of mesophase is reminiscent of charge density waves [11] or spin density waves in crystals [10]. Here we demonstrate a mechanical analog of a predominantly elastic character. Although we consider here such mesoscopic superstructures only for colloidal systems explicitly we expect that our findings should have a much wider applicability: related phenomena could occur for dusty plasmas [12], lattices of spherical block copolymer micelles under confinement [13], superstructures of small molecules or atoms adsorbed on stepped surfaces [14], etc. Boundary-induced strain fields are a commonly observed phenomenon, e.g. for quantum dot superlattices [15], lattice mismatched fused GaAs/InP wafers [16], $Ge_xSi_{(1-x)}$/Si heterostructures [17] and rotationally misaligned Si wafers covalently bonded together [18]. In the last case, a careful X-ray analysis did reveal a periodic displacement pattern, qualitatively similar to the findings we shall report below. However, while the experimental observation of strain distributions on atomic scales (via electron microscopy or x-ray scattering) is difficult, elastic distortions and lattice defects can be very nicely visualized in colloidal crystals [19].

**Model and simulation protocol**

We describe the colloidal particles as point particles interacting with a potential $V(r) = \varepsilon(\sigma/r)^{12}$, where $\varepsilon$ sets the energy scale and $r$ denotes the interparticle distance. As is well known, at low enough temperatures $T$ this system forms a crystal with a triangular lattice structure, where the lattice parameter $a$ is related to the chosen density $\rho$ via $a^2 = 2/(\sqrt{3}\rho)$. Of course, for such systems with inverse power law potentials, $T$ and $\rho$ are not independent control parameters, in our case it is only the combination $\rho(\varepsilon / k_B T)^{1/6}$ that matters [20]. Thus, choosing length units such that $\rho = 1.05$ one finds that the (presumably continuous) melting transition of the crystal occurs at [21] $k_B T_m / \varepsilon \approx 1.35$. In the present work, however, we are not concerned with the melting behavior of the system, and shall consider only temperatures $k_B T / \varepsilon = 1$ or below. Note that due to the finite width D of the confined crystals that we consider the melting transition is strongly smeared out (already in the fluid phase surface-induced crystalline layers occur close to the walls [9]).

We here use a $1/r^{12}$ potential rather than the experimentally realizable [22] $1/r^3$ potential since we expect that the qualitative character of the phenomena under study is independent of the range of the potential, and moreover the present faster decaying potential is computationally much more convenient. Furthermore the present choice has been used in related earlier work [9, 21].

Following Ricci et al. [9], we create a confinement potential commensurate with this lattice structure by putting two rows of frozen particle at either side of the crystal, which in our case has $n_x$ rows containing $n_y$ particles each, so that the crystal in the case where there is no misfit has the linear dimensions $L_y = n_y a$ and $D = n_x a \sqrt{3}/2$. However, choosing a smaller distance between the frozen rows on both sides of this crystalline strip we can enforce a misfit, such that $D = (n_x-\Delta)a \sqrt{3}/2$.

Choosing $n_x = 30$, $n_y = 108$ (and periodic boundary conditions in the y-direction parallel to the walls) we first choose the positions of the wall atoms such that $\Delta = 0$, with an initial condition of a perfect triangular lattice structure, and equilibrate the system at $k_B T / \varepsilon = 1$ by standard Monte Carlo methods [23]. Then we move the rows with the wall atoms closer to each other, in order to create a misfit with $\Delta = 0.25$, re-equilibrate the system (typical equilibration times were of the order of 8 million Monte Carlo steps per particle), increase the misfit to $\Delta = 0.5$, and so forth. Recording the stress $\sigma = \sigma_{yy} - \sigma_{xx}$ [24] from the computation of the virial tensor [20] we observe an almost linear increase up to a maximum value which is reached at $\Delta = 2.0$, where an abrupt first-order like transition to a slightly negative value of $\sigma$ is observed (Fig. 1). Further increase of $\Delta$ leads to an increase of $\sigma$ again.

Examination of the structure reveals that this sudden decrease of $\sigma$ is caused by a transition in the number of rows, $n_x \rightarrow n_x-1$. Since the particle number in our simulation is conserved, the $n_y$ extra particles of the row that has disappeared need to be redistributed in the system. These extra particles are not distributed at random, however, but in an almost regular way, giving rise to an almost periodic stress (and strain) pattern, see Fig. 2a.

Since both boundaries of the system are fully equivalent, one would expect that at both walls the deformation pattern is the same (apart form a translation in *y* direction). This, however, is not the case. We interpret this lack of symmetry by the problem that the defects that the extra particles cause in the original triangular lattice structure leads to the creation of many metastable states with similar (free) energies, which are only slightly higher than the true free energy minimum.

Thus we have attempted to guess the optimal structure by distributing an equal number of excess atoms, namely $n_d = n_y/(n_x-3)$, (= 4 for the present choice of $n_x$ and $n_y$) into each of the available $n_x-3$ rows, at randomly chosen positions in these rows, and then carefully equilibrate the system. Note that we observe (Fig. 2a) that the rows adjacent to the rows of fixed particles forming the boundaries stay free from excess particles, and thus still contain $n_y$ particles commensurate with the corrugation of the

``walls'', while all the $n_x$-3 inner rows then contain $n_y + n_d$ particles.

Indeed such a preparation of the systems leads to more regular pattern of local stresses and strains, in the form of standing waves (Fig. 2b). When we reduce $\Delta$ we find that this structure makes the inverse transition $n_{x-1} \rightarrow n_x$ at about $\Delta = 1.5$, indicative of some hysteresis, always to be expected for discontinuous transitions. However, on further reduction of $\Delta$ the data for $\sigma$ in the commensurate state with $n_x = 30$ rows are reproduced, within statistical errors, irrespective on the starting condition. In contrast, the parts of the curves for $\Delta \geq 2$ do not agree with each other, since the system that is incommensurate with the corrugated walls is easily locked into metastable states. However, we have tried various other ways of distributing excess particles (leaving two rows adjacent to the walls free of them, etc.), but we did not find stable states in this way, giving a strong hint that with the structure of Fig. 2b the free energy minimum has been correctly guessed.

**Soliton staircases and their characterization**

In order to study the structure of the incommensurate strip in more detail, Fig. 3 presents superimposed snapshot positions of the particles. The size of the superimposed irregular black dots thus shows roughly the typical mean square displacement of the particles. One can see that less displacement occurs in the rows adjacent to the row of fixed particles, and the arrangement of the particles is commensurate with the boundary. In the next rows, however, there are regions where the particles are in well-defined positions compatible with the commensurate row, while in between there are regions where the particle positions are almost fully smeared out along the y-direction. These regions represent solitons, where an extra particle needs to be accommodated, Fig.3b, and these solitons form a regular lattice in the y-direction. However, unlike standard solitons considered in the literature [25], these defects are spread out also in the x-direction perpendicular to the walls, and hence are not strictly one-dimensional objects.

Fig.4a shows that the strain density wave pattern is maintained when one considers lower temperatures, and Fig.4b shows that indeed the wavelength of the pattern is controlled by the thickness $D$ of the strip: for $n_x = 20$ the rule that $n_d = n_y/(n_x - 3) = n_y/17$ yields $n_d = 6$ for $n_y = 102$, and indeed 6 solitons can be identified at each boundary. Finally, Fig.5 characterizes the soliton staircase in terms of the displacement variable of the particles in a row from their original lattice position and their local fluctuations. The periodicity expected for a soliton staircase indeed is rather well developed. Choosing analogous data for rows further inward the picture is qualitatively similar, but the amplitude of the periodic variation is smaller.

While in the state where the transition $n_x \to n_{x-1}$ has not yet occurred a crystalline structure commensurate with the wall potential occurs, and hence long range order is enforced by the boundaries even for $L_y \to \infty$, this is not so for the incommensurate phase with the soliton lattices at the walls. Thus we expect that for $L_y \to \infty$ the soliton "lattice" is not strictly periodic, since the mean square displacements between subsequent solitons will increase linearly with the distance between the solitons in the thermodynamic limits at nonzero temperature. For finite $L_y$, however, the periodic boundary condition enforces a periodic soliton arrangement, on average (but this lattice then is not strictly pinned but can diffuse along the strip as a whole, due to translational invariance. On the time-scale of the Monte Carlo simulation at $T = 0.1$, we have not observed a significant diffusion of this type, however).

Thus understanding qualitative aspects of the soliton lattice (such unstable widths of the peaks in Fig. 5b) is a subtle matter: we hope to analyze this issue in a more detailed publication. Note, however, that real systems are finite also in the y-direction, and may also have some confining walls rather than a periodic boundary condition, so this lack of order for $L_y \to \infty$ may be a somewhat academic question.

Of course, the fact that the boundary potential in our model is so strong that the rows $k=1$ and $k= n_x -1$ that are immediately adjacent to the fixed particles defining the boundaries stay commensurate is an unimportant detail of our model. It is clear that in the physical realizations of confinement of colloidal crystals e.g. via laser fields or other experimental means providing particle confinement such features need not carry over. Thus, the precise relation between the period $\ell$ of the soliton staircases and the confining width *D* will be somewhat model-dependent, but we believe that the general mechanism of boundary-induced periodic strain patterns demonstrated here should be widely applicable.

**Concluding remarks**

By simulations of a generic model we have discovered a new state of matter where due to confining boundaries in a two-dimensional crystalline strip a long-wavelength periodic strain field pattern is realized, the case of the elastic distortions being given by the steps of the soliton staircase close to the boundaries. Of course, it has been well known that confining a two-dimensional crystal between two parallel boundaries a distance *D* apart invariably creates some misfit, except for ``magic numbers'' of *D* where the misfit is zero. For small $n_x$ such situations have been studied in the literature [24,26,27,28,29]. However, the new feature of the present work is the study of long-wavelength boundary-induced superstructures that occur when $n_x$ is large. We have described these super-structures as standing strain waves caused by the soliton lattices that form along the boundaries.

We expect that closely related phenomena are experimentally realizable not only in colloidal crystals and various other soft matter systems, but also for epitaxial growth on striped crystalline substrates, semiconductor heterostructures, and fused wafers. Our findings may provide interesting perspectives to the field of strain engineering [30,31,32,33] for different potential applications.

**Acknowledgements**

This research was partially supported by the Deutsche Forschungsgemeinschaft, Project TR6/C4. We are grateful to P. Nielaba, A. Ricci and I. Snook for very useful interactions.

**Figure Caption**

Figure 1: Internal stress $\sigma = \sigma_{yy} - \sigma_{xx}$ (in LJ units) in the confined crystalline strip plotted vs. $\Delta$, for the case of a system started with $n_x = 30$, $n_y = 108$ (full symbols) and a system started with $n_x = 29$, $n_y = 108$ plus the appropriate extra particles, as described in the text. Curves are guides to the eye only.

Upper insert shows a schematic sketch of our geometry: we study a system of size D in x-direction and $L_y$ in y-direction, apply periodic boundary condition along the y-axis, while the boundary in the x-direction is created by two-rows of fixed particles (shaded) on the ideal positions of a perfect triangular lattice at each side. In the fully commensurate case, $D = (n_x-\Delta) a \sqrt{3}/2$.

Figure 2: Strain patterns of a system at $\Delta = 2.0$, started with $n_x = 30$, $n_y = 108$, (a) after the transition $n_x \to n_x-1$ has occurred and (b) of a systems started with $n_x = 29$, $n_y = 108$ plus extra particles, as described in the text. The strain is calculated from the particle configurations, $u_{ij} = \partial u_i / \partial x_j + \partial u_j / \partial x_i$, where $u_i(\vec{R}_n)$ is the $i$-component of the displacement vector of the particle labelled by $n$ relative to the site $\vec{R}_n$ of the reference lattice, $\{\vec{x}_i\}$ are the (two-dimensional) cartesian coordinates $(x,y)$, an averaging is performed over 1000 cartesian coordinates (x,y) and an averaging is performed over 1000 configurations separated by 100 Monte Carlo steps in the course

of the simulation. Note that the numbers shown along x- and y-axis indicate the Cartesian coordinates of the particles, and not the particle row numbers.

Figure 3: (a) Particle configuration in a system with $n_x = 30$, $n_y = 108$, $\Delta = -2.6$, where a transition $n_x \to n_x-1$ has occurred, at a temperature $k_B T / \varepsilon = 1$. The figure shows 750 superimposed positions of the particles. (b) Close-up of the structure near the upper wall. Numbers shown along the axes indicate the cartesian coordinates of the particles.

Figure 4: Strain patterns of a system of the type as shown in Fig. 1 but for (a) $n_x = 108$, $n_y = 30$, $k_B T / \varepsilon = 0.1$ and (b) $n_x = 108$, $n_y = 20$, $k_B T / \varepsilon = 0.1$. The calibration bars are shown on the right hand side of the graphs.

Figure 5: (a) Variation of the phase variable defined as $<u_x(\vec{R}_n)> - X_n$ versus the index of the x-coordinate, for the rows $k = 2$ and 28 of a system $n_x = 108$, $n_y = 30$, $k_B T / \varepsilon = 0.1$ and (b) Corresponding Lindemann parameters (l(n)) defined as $<u_x^2(\vec{R}_n)> - <u_x(\vec{R}_n)>^2$.

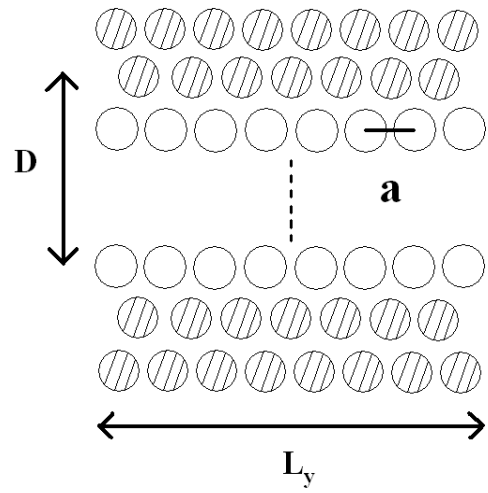

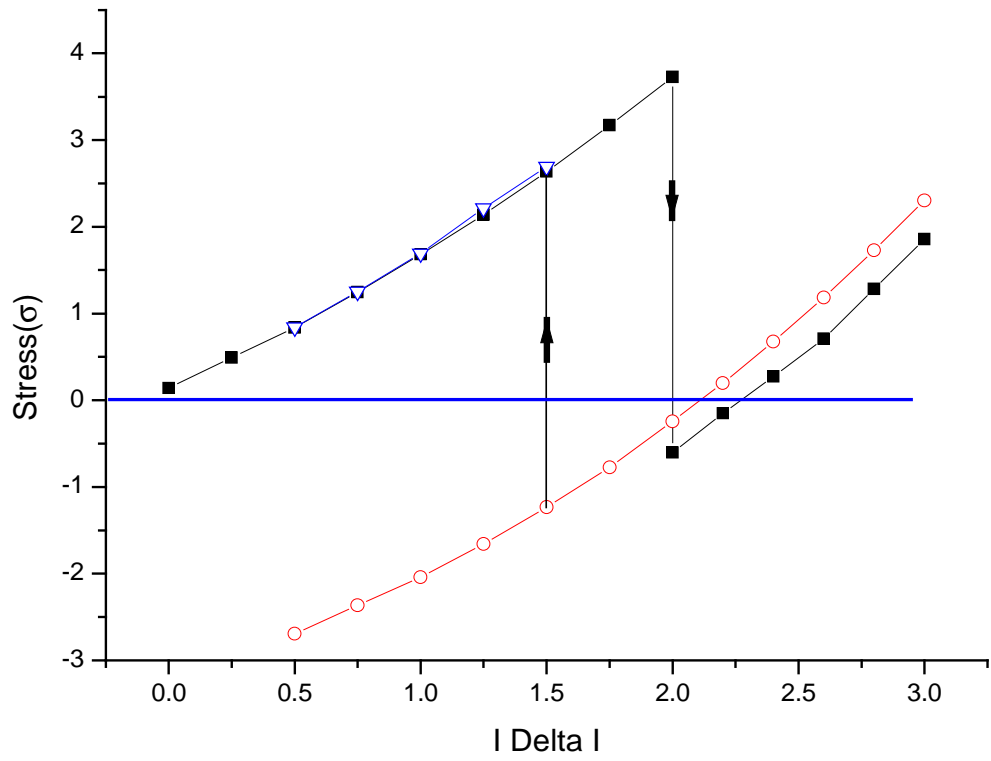

**Figure 1**

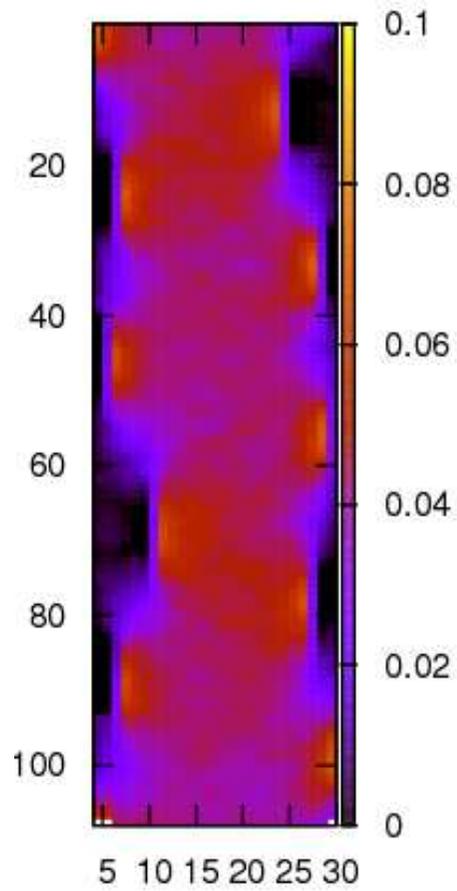 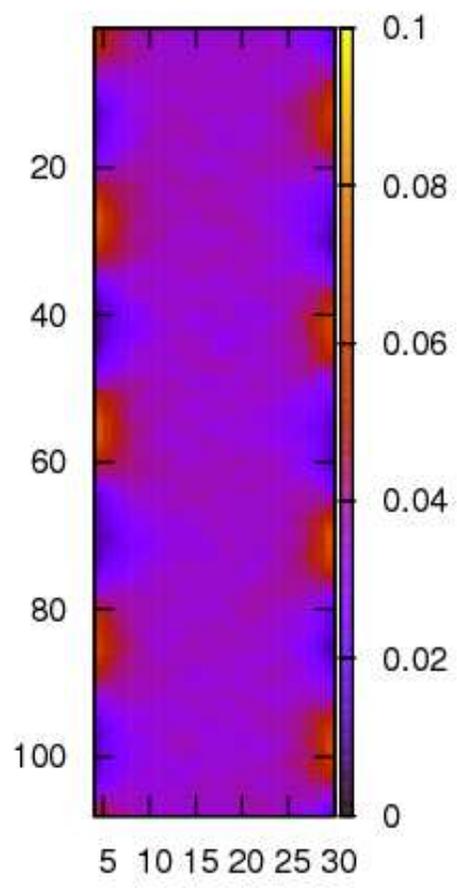 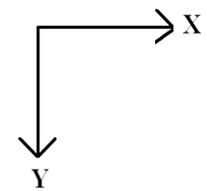

(a) (b) **Figure 2**

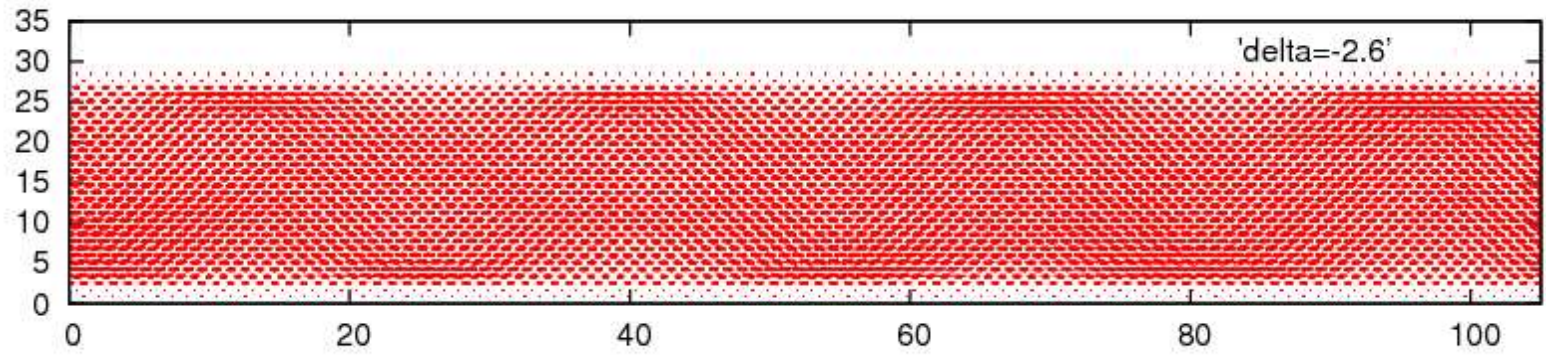

(a)

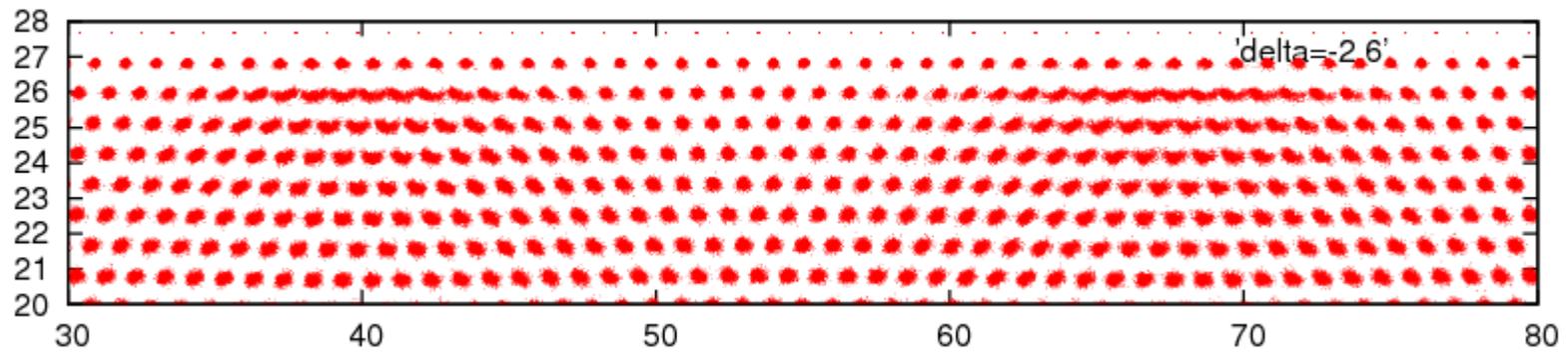

(b)

**Figure 3**

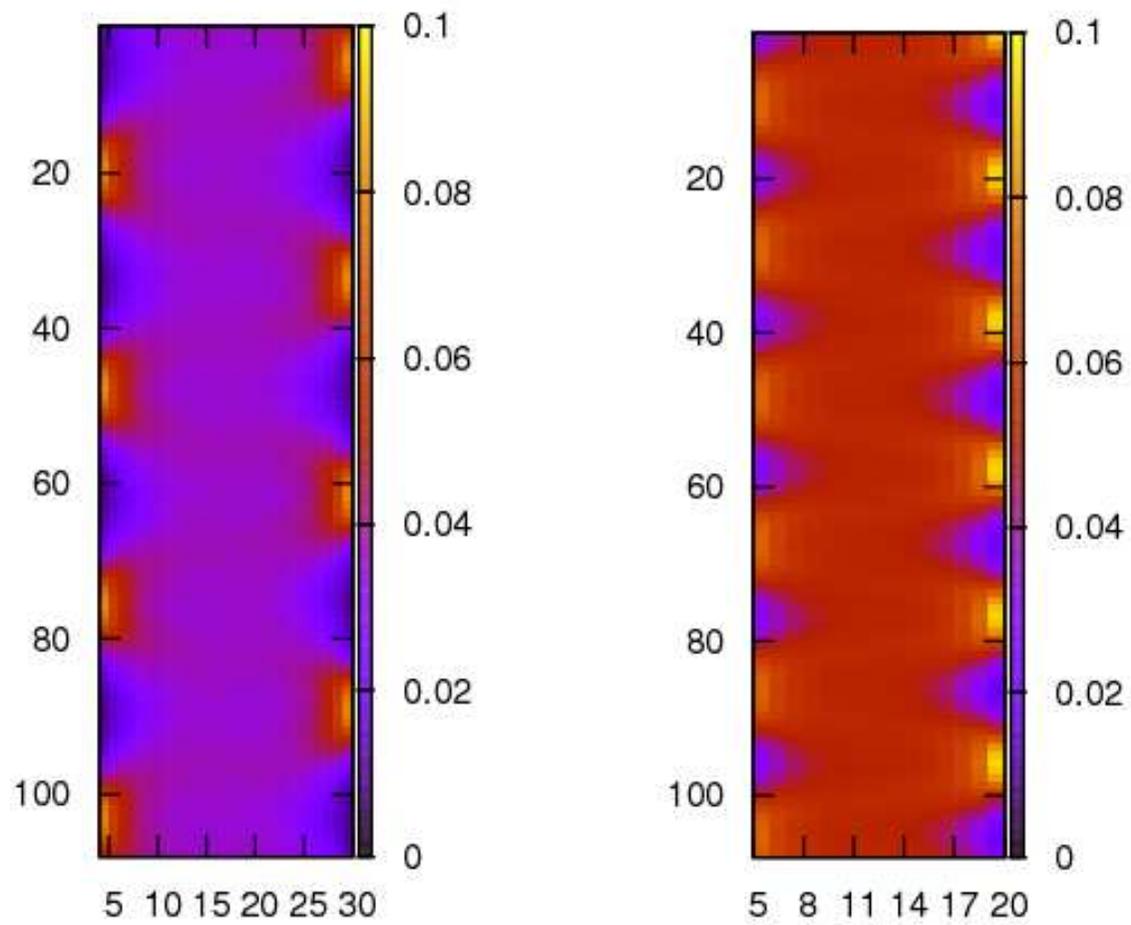

(a)          (b)          **Figure 4**

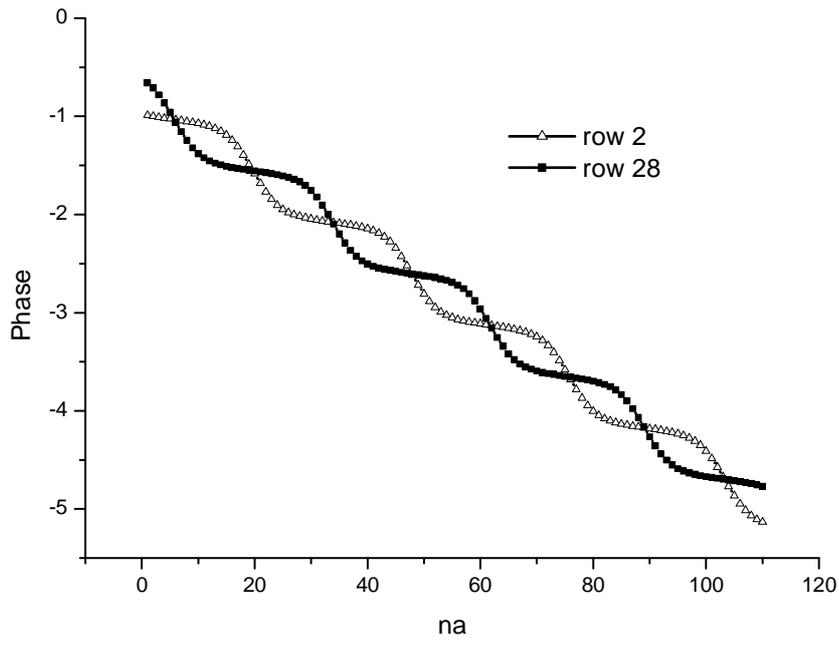

(a)

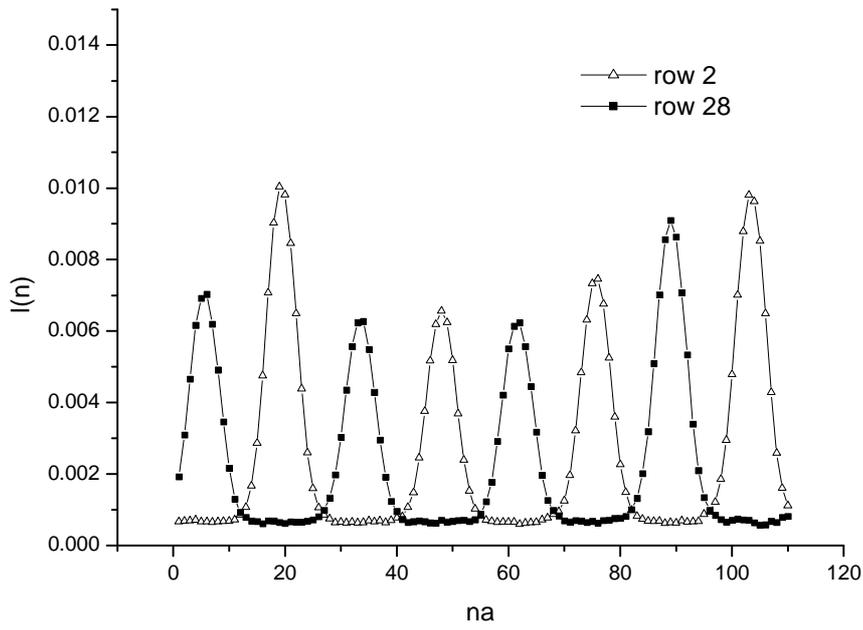

(b)

**Figure 5**